# Bonding and anti-bonding modes in metal-dielectric-metal plasmonic antennas for dual-band applications

*Kateryna L. Domina[1], Vyacheslav V. Khardikov[1,2], Vitaliy Goryashko[3*] and Alexey Y. Nikitin[4,5,6*]*

[1]School of Radio Physics, V. N. Karazin Kharkiv National University, 4, Svobody Square, Kharkiv 61022, Ukraine.
[2]Institute of Radio Astronomy of National Academy of Sciences of Ukraine, 4, Mystetstv Street, Kharkiv 61002, Ukraine.
[3]FREIA Laboratory, Uppsala University, Uppsala, Sweden.
[4]Donostia International Physics Center (DIPC), 20018 Donostia-San Sebastián, Spain.
[5]IKERBASQUE, Basque Foundation for Science, 48013 Bilbao, Spain.
[6]CIC nanoGUNE, 20018, Donostia-San Sebastián, Spain.

*Correspondence to: vitaliy.goryashko@physics.uu.se, alexey@dipc.org

**Resonant optical antennas supporting plasmon polaritons (SPPs) – collective excitations of electrons coupled to electromagnetic fields in a medium – are relevant to sensing, photovoltaics and light emitting devices, among others. Due to the SPP dispersion, a conventional antenna of fixed geometry, exhibiting a narrow SPP resonance, cannot simultaneously operate in two different spectral bands. In contrast, this study demonstrates that in metallic disks, separated by a nanometric spacer, the hybridized anti-bonding SPP mode stays in the visible range, while the bonding one can be pushed down to the mid-infrared range. Such an SPP dimer can sense two materials of nanoscale volumes, whose fingerprint central frequencies differ by a factor of 5. Additionally, the mid-infrared SPP resonance can be tuned by employing a phase-change material ($VO_2$) as a spacer. The dielectric constant of the phase-change material is controlled by heating the material at the frequency of the anti-bonding optical mode. Our findings open the door to a new class of optoelectronic devices able to operate in significantly different frequency ranges in the linear regime, and with the same polarization of the illuminating wave.**

Optical antennas are used to couple freely propagating optical fields to fields in matter and already present an established concept in nanophotonics.[1, 2] They can be used for nanoscale imaging, light emission, optical detection, coherent light control, as well as for life sciences and medical applications.[3, 4] Optical antennas are typically made of metals supporting SPPs (plasmonic antennas)[1] and can be designed for a very broad frequency range, from visible (VIS) to terahertz frequencies. At mid-infrared (mid-IR) frequencies, novel two-dimensional and van der Waals materials have been suggested as an alternative for optical antennas, as for instance graphene[5, 6] and topological insulators[7] (supporting Dirac plasmon polaritons) or h-BN (supporting hyperbolic phonon polaritons).[8, 9] The latter materials, on the one hand, can be more interesting compared to metals due to both much higher polariton confinement and active tuning of polariton wavelength (and thus antenna resonance). On the other hand, they are very limited to relatively narrow spectral intervals, in which they support polaritons.

A particularly interesting application, for which optical antennas present a powerful tool, is surface-enhanced spectroscopy.[10] Strong electromagnetic near-fields created by the antennas can dramatically (by orders of magnitude) enhance the absorption resonances of the placed nearby molecules. The molecular fingerprints are recognized either by analyzing Fano-like features appearing in the SPP resonances in the spectra of the extinction cross-section of the antennas (surface-enhanced absorption spectroscopy[10]) or by monitoring Raman signals (surface-enhanced Raman scattering[11]). Surface-enhanced spectroscopy is used both in the mid-IR frequency range, in which many organic materials have vibrational resonances,[10, 12, 13] and in the visible one, where electronic transitions in molecules are more relevant.[14, 15] Since metallic antennas exhibit SPP resonances in a broad frequency range (with the resonance frequency depending upon antenna's geometry), they seem to be more universal for surface-enhanced spectroscopy applications, than antennas made of alternative materials. At the same time, due to a big difference between SPP wavelengths at VIS and mid-IR frequencies, one and the same conventional SPP antenna cannot simultaneously operate in both spectral bands. For enabling dual-band performance, one has to design antennas with large geometric aspect ratios (e.g. rod antenna).[16] In this case, the antenna responds differently to illuminating waves of different in-plane polarizations.

Here we show that by using a simple vertical SPP dimer composed of two metallic disks separated by a nanometric spacer, it is possible to achieve well-pronounced SPP resonances simultaneously in the VIS and mid-IR ranges. By numerical simulations, we demonstrate that each of the resonances can be used for surface-enhanced spectroscopy. The frequencies of the emerging SPP resonances are predominantly defined by strong hybridization between the SPPs supported by each disk. Therefore, the resonances of the vertical SPP dimer can be tuned by the thickness of the spacer and its dielectric permittivity. As an example, we tune the position of the mid-IR resonance by filling the spacer of the antenna with $VO_2$ (phase-change material having temperature-dependent dielectric permittivity). Importantly, thanks to the rotational symmetry of the antenna, antenna's electromagnetic response is totally independent of the in-plane polarization of the incident wave.

In Figure 1a, we illustrate the concept of a dual-band antenna: two gold disks (of height $h = 40$ nm and radius $R = 120$ nm) are separated by a dielectric spacer. The antenna is placed on a standard $SiO_2$ substrate, transparent in both VIS and mid-IR ranges. For simplicity of simulations, we consider a periodic array of such antennas. Taking a subwavelength period, $L = 400$ nm, in the whole studied range of wavelengths (from 0.55 μm to 5.5 μm) we avoid diffraction into higher orders except the fundamental (zero) one. On the other hand, the chosen period is much larger than the characteristic confinement length of the SPPs in the disks, $L \gg \delta_{SPP}$, where $\delta_{SPP} \lesssim \lambda_p/2\pi$ (with $\lambda_p$ being the wavelength of the SPP on an infinite metal surface). Hence, the coupling between the SPP modes of different antennas is negligible.

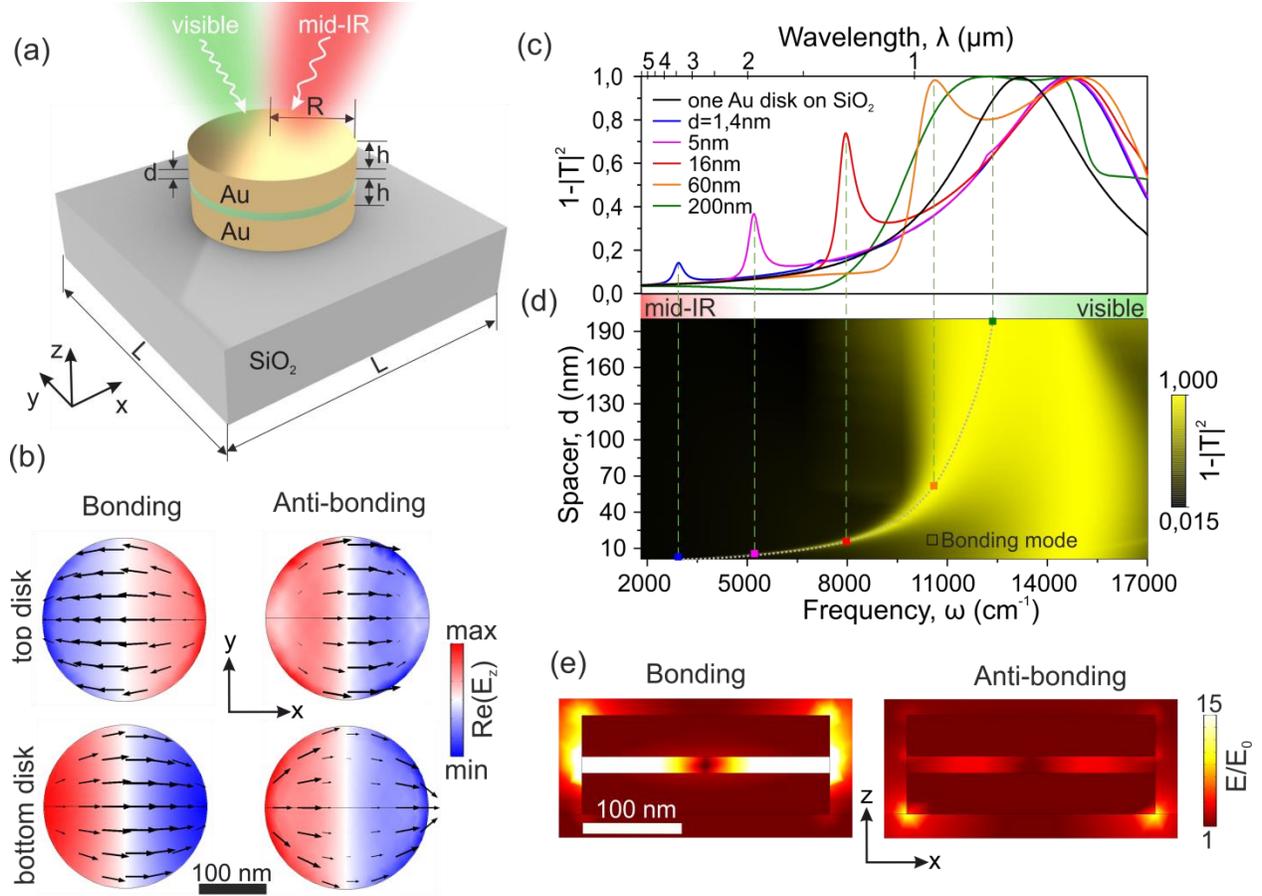

**Figure 1**. Bonding and anti-bonding modes in a vertical SPP dimer antenna placed on $SiO_2$ substrate. a) The schematic of the antenna. The antenna is illuminated by a normally incident electromagnetic plane wave with the polarization along the $x$-axis. b) Spatial distribution of $Re(E_z)$ taken on the bottom surface of the upper disk (upper images) and top surface of the lower disk (lower images), respectively. The arrows represent the in-plane electric field c) The extinction spectra of the periodic array of antennas with different spacers between the disks. d) The extinction of the periodic array of antennas as a function of the wavelength (frequency) and the spacer thickness. The black dashed curve represents the position of the bonding resonance according to the coupled oscillators model. e) Spatial distribution of the absolute value of the electric field. The frequencies of the bonding and anti-bonding modes in (b) and (e) are 7937 and 14705 cm$^{-1}$, respectively, while the spacer is $d = 16$ nm. In all panels the radii of the disks are 120 nm. The period of the array is $L = 400$ nm.

The extinction (defined as $1 - |T|^2$, with $T$ being the amplitude transmission coefficient) spectra for different values of the spacer thickness, $d$, is shown in Figure 1c. As a reference, the extinction for a single gold disk is represented by the black curve, showing one dipolar SPP resonance around $\lambda = 760$ nm ($\omega = 13158$ cm$^{-1}$). When the antenna is composed of the two disks, the dipolar SPP modes of the upper and lower disks couple, building up the hybridized bonding and anti-bonding modes.[17] Accordingly, the extinction spectrum of the dimer manifests two peaks (colored curves in Figure 1c), with the distance between them being increased as $d$ decreases. While the anti-bonding (high-frequency) mode remains almost at the same spectral position (near $\omega = 14500$ cm$^{-1}$), the bonding (low-frequency) mode quickly moves away towards mid-IR frequencies. The effective repulsion of the two resonant peaks is more clearly visible in the colorplot representing the extinction as a function of both $\omega$ and $d$ (Figure 1d).

As it is clearly seen from Figure 1d, for $d > R$ the resonant wavelengths of the bonding and antibonding modes are symmetrically shifted to longer and shorter wavelengths, respectively, with respect to the resonant wavelength of a single disk. These shifts are the manifestation of the interaction between the two disks via the common dipole field, see the Supporting Information. It turns out that according to the quasistatic approximation (assuming that the spheroid is small compared to the resonant wavelength of the mode) the resonant wavelengths of the vertical dimer for $d > R$ are $\lambda_{\pm} \approx \lambda_r(1 \mp \beta/2)$. Here, $\lambda_r$ is the resonant wavelength of a single disk (S4) and $\beta$ is the coupling between the disks. This coupling is proportional to $(1 + d/w)^{-2}$ so that the separation between the resonant wavelengths increases as the distance between the disks $d$ decreases. The model is developed for $d > R$ but the same functional dependence on $d$ can be used for the bonding mode in a wider parameter space: $\lambda_B/\lambda_r = 1 + a/(1 + d/w)^2$. Here, $\lambda_B$ is the resonant wavelength of the bonding mode; $a$ is a constant defining the integrated spatial overlap of the dipolar modes and $w$ is the effective size of the overlap. Fitting the parametric dependence $\lambda_B(d)$ to the simulation data (Figure 1d, colorplot), we get the values $a = 4$ and $w = 8.33$ nm. The result of the fit is shown in Figure 1d, by the black dashed curve. Our simple expression for the wavelength of the bounding mode is useful for a quick estimate of the wavelength of the bonding mode prior to numerical simulations. Moreover, the dependence $\lambda_B(d)$ can be used for an order-of-magnitude estimates for disks of different radii since $w \sim R/10$ (for details, see the discussion about $w$ after (S6)).

Importantly, the strong dependence of the bonding resonance on the spacer thickness allows for its significant shift towards low frequencies, down to the mid-IR range. This dependence can be used for efficient tuning of the antenna, for example by applying strain or by thermally induced melting of a self-assembled layer placed inside the spacer.[18] The coupled oscillators model also qualitatively provides the spatial field distribution of the hybridized SPP modes: the effective electric in-plane dipoles induced in the upper and lower disks are either parallel (anti-bonding mode) or anti-parallel (bonding mode). This result is fully consistent with the in-plane electric field spatial distributions found by means of full-wave electromagnetic simulations (Figure 1b, black arrows distribution). The bonding and anti-bonding modes can also be easily recognized by the anti-symmetric and symmetric spatial distribution of the real part of the vertical component of the electric field, $E_z$ (being proportional to the surface charge density), on the top face of the lower disk and on the bottom face of the upper disk (colorplot in Figure 1b).

Due to the anti-symmetric in-plane electric field distribution (and thus the reduced effective in-plane dipole moment, preventing an efficient coupling with the incident wave), the peak intensity of the bonding mode is lower as compared to the anti-bonding one.[17] In a planar SPP dimer (two gold disks located in the same plane next to each other), one of the modes is always dark, *i.e.* it cannot be excited by a normally-incident light.[8] In contrast, for our vertical SPP dimer, the situation is different thanks to the retardation effect (phase shift) between the effective dipoles induced in the disks. Therefore, both modes can be excited by a normally-incident plane wave even in the case of the symmetric dielectric environment of the vertical SPP dimer. Apart from the different symmetry of the fields of the hybridized SPP modes, the spatial distribution of the electric field intensity, $|\boldsymbol{E}|$, is significantly different as well. Namely, while the maxima of $|\boldsymbol{E}|$ of the anti-bonding modes are concentrated around the edges of the bottom ring (right panel of Figure 1e), the $|\boldsymbol{E}|$ of the bonding mode is strongly "pushed" into the spacer, (left panel in Figure 1e and the inset). The spatial distribution of $|\boldsymbol{E}|$ provides a hint to the best spatial allocation for the sensed material. In particular, the spectral position of the bonding mode should be very sensitive to the dielectric permittivity of the spacer. The concept of probing minute amounts of materials inside the spacer has recently been suggested for similar "screened" plasmonic modes in graphene resonators above a metal pad [19] and later realized experimentally.[20, 21]

In order to test the sensing functionality of our dual-band SPP dimer antenna, we use two different materials: octane and j-aggregate, with their dielectric permittivities $\varepsilon_{\text{oct}}$ and $\varepsilon_j$ taken from[22] and,[15] respectively. While octane shows a typical for alkenes absorption resonance of the stretching vibration of the C-H bond at mid-IR frequencies (around $\omega = 3000$ cm$^{-1}$, see Figure 2a), j-aggregate has an excitonic absorption peak in the visible range (around $\omega = 14500$ cm$^{-1}$, see Figure 2b). For simplicity of the proof of principle, we consider cylindrically-symmetric shapes of the probed materials, preserving the symmetry of the antenna. According to the spatial positions of the $|\boldsymbol{E}|$ maxima (Figure 1e), we place 1.4 nm-thick layer of octane (in the form of a disk) between the gold disks and 3-nm thick ring of j-aggregate around the bottom gold disk, as shown in the insets to Figure 2c. The octane disk radius coincides with that of the gold rings, so that its volume is 63.3x10$^3$ nm$^3$ (i.e. 3.9x10$^{-19}$ mol) and thus fits 2.35x10$^5$ octane molecules. In contrast, the j-aggregate ring has the internal radius coinciding with the radius of the gold rings and a width of 3 nm (so that its volume is 6.9x10$^3$ nm$^3$), and mimics a single j-aggregate rod (typically found in solutions), rolled around the antenna. Figure 2c shows the extinction spectra of the SPP dimer antenna with the molecular layers (red solid curve) and with the reference layers having the constant dielectric permittivities $\varepsilon_1$ and $\varepsilon_2$ (black dashed curves). The values of $\varepsilon_1$ and $\varepsilon_2$ correspond to the average of $\varepsilon_{\text{oct}}$ and $\varepsilon_j$, respectively, in the shown frequency range. The comparison between the spectra clearly reveals the absorption fingerprints in both VIS and mid-IR spectral bands. Remarkably, the visibility of the absorption fingerprints exceeds the absorption by the octane disk and j-aggregate ring by about two orders of magnitude (see orange and violet curves in Figure 2c). Interestingly, apart from the excitonic resonance (approximately repeating the peak of Im($\varepsilon_j$)) the absorption by the j-aggregate ring also shows a broader exciton-polaritonic resonance at higher frequencies, where Re($\varepsilon_j$) takes negative values.[23, 24] The observed significant fingerprint enhancement justifies the use of our dual-band dimer SPP antenna for surface-enhanced spectroscopy. Notice that from the practical point of view, the delivery of the sensed materials in between the gold disks, as well as around the antenna, can potentially be realized via nanofluidics.[25]

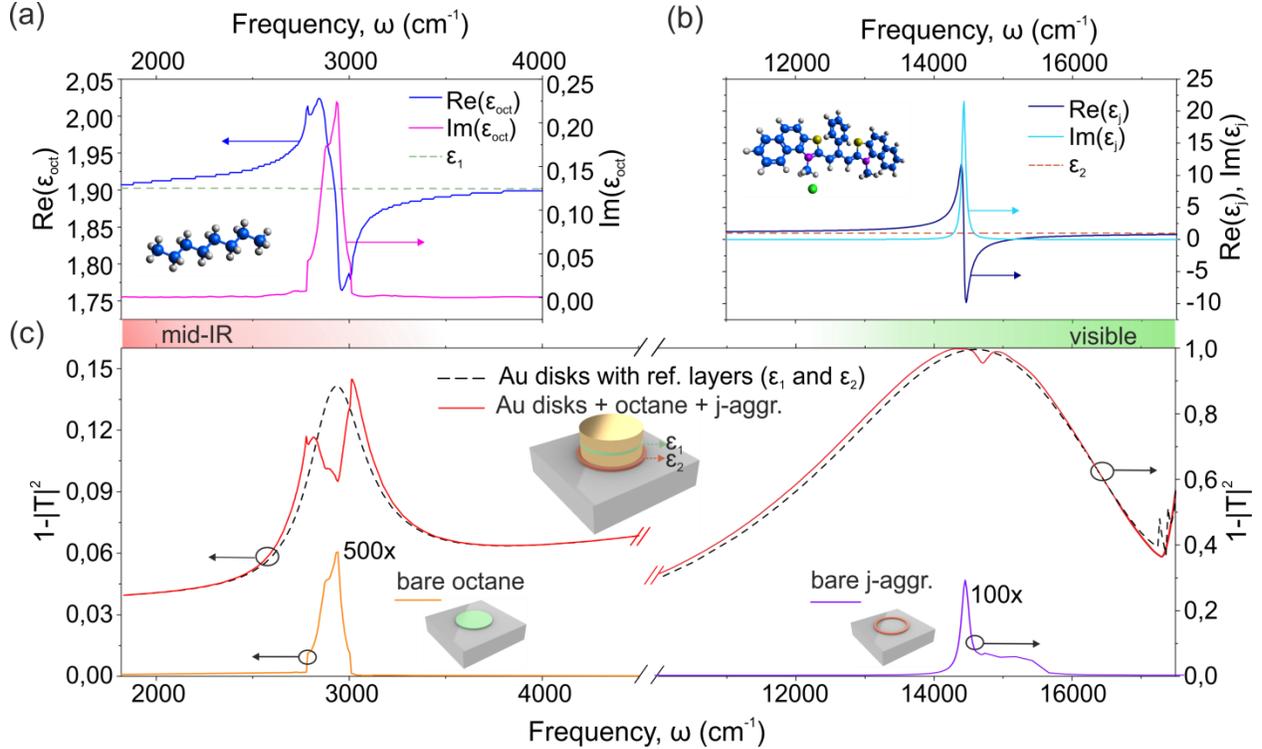

**Figure 2.** Surface-enhanced dual-band spectroscopy with the SPP dimer antenna. a,b) Real and imaginary parts of the dielectric permittivity of octane (a) and j-aggregate (b), as a function of frequency. The horizontal dashed lines mark the value of the reference dielectric permittivities. The insets show the atomic structure of the molecules. c) Extinction spectra of the SPP dimer antenna with the molecules (red solid curve) and with the molecular dielectric permittivities replaced by the reference ones (black dashed curve). Orange and violet curves represent the absorption spectra of the octane disk and j-aggregate ring, respectively. The parameters of the gold disks are the same as in Figure 1.

The strong field confinement inside the spacer of the SPP dimer can enable the spectral tuning of the bonding resonance by changing the dielectric permittivity of the spacer. An interesting possibility to control the dielectric permittivity is via heating of phase change materials, such as, for instance, $VO_2$. At the rutile to monoclinic transition temperature ($T_M = 67$ °C), $VO_2$ exhibits a metal to semiconductor transition (Mott transition) in its electronic structure: the rutile phase is metallic while the monoclinic phase is semiconducting.[26] As a result, its dielectric permittivity, $\varepsilon_{VO2}$, strongly changes with the temperature, $T$, around $T = T_M$, particularly in the mid-IR frequency range (Figure 3a).[27, 28] For this reason, $VO_2$ has recently been used for active switching of SPPs[29] and reconfigurable control of in-plane phonon polariton propagation.[30] To demonstrate the tunability of our SPP dimer antenna, we perform the simulation of its extinction spectra, placing $VO_2$ between the gold disks (see Figure 3b). As seen in the figure, in the temperature range 50-60 °C (i.e. heating up the antenna by 10 °C) the mid-IR resonant peak blueshifts by 240 wavenumbers. The blueshift of the resonance with $T$ can be explained by the reduction of the real part of $\varepsilon_{VO2}$ (Figure 3a, solid curves). On the other hand, as the imaginary of $\varepsilon_{VO2}$ increases with $T$ (Figure 3a, dashed curves), the amplitude of the mid-IR peak decreases. From the results presented in Figure 3, we can directly estimate the figure of merit of the bounding resonance as FOM = $\Delta\lambda/\Delta n \cdot$ FWHM $\approx 4$, where $\Delta\lambda$ is the wavelength shift per refractive index change, $\Delta n$, and FWHM is the full width at half maximum. In contrast, the anti-bonding resonance at the VIS frequencies does not display any significant change with $T$. Thus, the anti-bonding SPP resonance in the VIS range can be used for controlling the bonding SPP resonance in the mid-IR range by means of optical pumping.[31] Such pumping can be practically realized with a basic, mW-scale, commercial diode laser. For instance, for a thin silicon substrate, the heat flux across the face 1 mm in height and 1 cm in width amounts to less than 1 mW. The main expected mechanism of thermal losses is due to blackbody radiation, which for a sample temperature of 40 degrees above the ambient, results in 30 mW of losses. Notice that previously SPP resonances in plasmonic antennas have already been explored for the experimental demonstration of antenna-assisted, optically-triggered phase transition of $VO_2$ at the near-IR frequencies.[32]

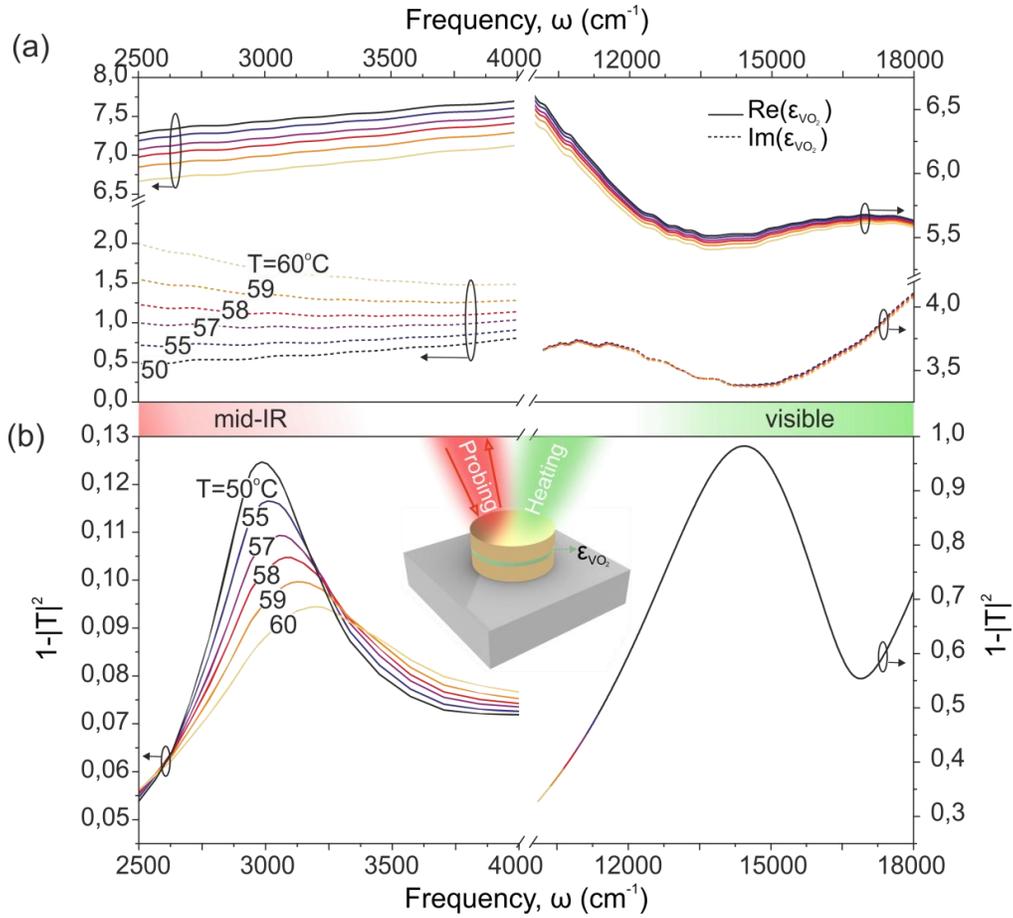

**Figure 3**. Tuning of the mid-IR resonance by heating $VO_2$. a) Real part (solid curves) and imaginary part (dashed curves) of the dielectric permittivity of $VO_2$, $\varepsilon_{VO2}$, as a function of $\omega$, at different temperatures (the data is taken from.[28] b) The extinction spectra of the SPP dimer antenna at different temperatures (close to Mott transition) in the VIS and mid-IR ranges. The parameters of the gold disks are the same as in Figures 1,2 and $VO_2$ spacer is $d = 6$ nm.

In conclusion, we have suggested a simple dual-band vertical dimer antenna exhibiting SPP resonances simultaneously in the VIS and mid-IR frequency ranges. We have shown the tuning of the mid-IR resonance by means of heating the phase-change material (VO$_2$) placed in the spacer between the gold disks. The heating of VO$_2$ can be realized by resonant pumping of the anti-bonding SPP mode with a mW-scale, visible laser. We have demonstrated the potential of the SPP dimer antenna for dual-band surface enhanced spectroscopy. Although we have illustrated the identification of the fingerprints of two different molecules, such dual-band spectroscopy can also be used for identification of the same material (molecules) having its fingerprints in the VIS and mid-IR frequency regions. Moreover, with proper matching of the oscillator strength of the SPP resonances to that of the molecular absorption resonances, one can potentially achieve strong coupling between the SPPs of the antenna and the molecule, simultaneously in the VIS and mid-IR ranges. From a different perspective, our findings can pave the way to enhancing the efficiency of the Förster resonance energy transfer or photon upconversion applications (e.g. solar spectrum conversion for photovoltaics).

**Acknowledgements**
The authors acknowledge financial support from STINT Initiating Grant (IB2017-7026), A.Y.N. acknowledges the Spanish Ministry of Economy, Industry, and Competitiveness (national project MAT2017-88358-C3-3-R).

**Conflict of Interest**
The authors declare no conflict of interest.

[1]     P. Bharadwaj, B. Deutsch, L. Novotny, *Adv. Opt. Photon.* **2009**, *1*, 438.
[2]     P. Mühlschlegel, H.-J. Eisler, O. J. F. Martin, B. Hecht, D. W. Pohl, *Science* **2005**, *308*, 1607.
[3]     J. Alda, J. M. Rico-García, J. M. López-Alonso, G. Boreman, *Nanotechnology* **2005**, *16*, S230.
[4]     H. Xin, B. Namgung, L. P. Lee, *Nature Reviews Materials* **2018**, *3*, 228.
[5]     Z. Fang, Y. Wang, A. E. Schlather, Z. Liu, P. M. Ajayan, F. J. García de Abajo, P. Nordlander, X. Zhu, N. J. Halas, *Nano Letters* **2014**, *14*, 299.
[6]     A. Y. Nikitin, P. Alonso-González, S. Vélez, S. Mastel, A. Centeno, A. Pesquera, A. Zurutuza, F. Casanova, L. E. Hueso, F. H. L. Koppens, R. Hillenbrand, *Nature Photonics* **2016**, *10*, 239.
[7]     M. A. Poyli, M. Hrtoň, I. A. Nechaev, A. Y. Nikitin, P. M. Echenique, V. M. Silkin, J. Aizpurua, R. Esteban, *Physical Review B* **2018**, *97*, 115420.
[8]     J. D. Caldwell, A. V. Kretinin, Y. Chen, V. Giannini, M. M. Fogler, Y. Francescato, C. T. Ellis, J. G. Tischler, C. R. Woods, A. J. Giles, M. Hong, K. Watanabe, T. Taniguchi, S. A. Maier, K. S. Novoselov, *Nature Communications* **2014**, *5*, 5221.

[9]     F. J. Alfaro-Mozaz, P. Alonso-González, S. Vélez, I. Dolado, M. Autore, S. Mastel, F. Casanova, L. E. Hueso, P. Li, A. Y. Nikitin, R. Hillenbrand, *Nature Communications* **2017**, *8*, 15624.
[10]    F. Neubrech, C. Huck, K. Weber, A. Pucci, H. J. C. r. Giessen, *Chemical reviews* **2017**, *117*, 5110.
[11]    S. J. A. C. I. E. Schlücker, *Angewandte Chemie International Edition* **2014**, *53*, 4756.
[12]    M. Autore, P. Li, I. Dolado, F. J. Alfaro-Mozaz, R. Esteban, A. Atxabal, F. Casanova, L. E. Hueso, P. Alonso-González, J. Aizpurua, A. Y. Nikitin, S. Vélez, R. Hillenbrand, *Light: Science &Amp; Applications* **2018**, *7*, 17172.
[13]    D. Rodrigo, O. Limaj, D. Janner, D. Etezadi, F. J. García de Abajo, V. Pruneri, H. Altug, *Science* **2015**, *349*, 165.
[14]    Y. B. Zheng, B. K. Juluri, L. Lin Jensen, D. Ahmed, M. Lu, L. Jensen, T. J. Huang, *Advanced Materials* **2010**, *22*, 3603.
[15]    N. T. Fofang, T.-H. Park, O. Neumann, N. A. Mirin, P. Nordlander, N. J. Halas, *Nano Letters* **2008**, *8*, 3481.
[16]    C. D'Andrea, J. Bochterle, A. Toma, C. Huck, F. Neubrech, E. Messina, B. Fazio, O. M. Maragò, E. Di Fabrizio, M. Lamy de La Chapelle, P. G. Gucciardi, A. Pucci, *ACS Nano* **2013**, *7*, 3522.
[17]    Y.-C. Chang, S.-M. Wang, H.-C. Chung, C.-B. Tseng, S.-H. Chang, *ACS Nano* **2012**, *6*, 3390.
[18]    G. Hajisalem, M. S. Nezami, R. Gordon, *Nano Letters* **2017**, *17*, 2940.
[19]    S. Chen, M. Autore, J. Li, P. Li, P. Alonso-Gonzalez, Z. Yang, L. Martin-Moreno, R. Hillenbrand, A. Y. Nikitin, *ACS Photonics* **2017**, *4*, 3089.
[20]    I.-H. Lee, D. Yoo, P. Avouris, T. Low, S.-H. Oh, *Nature Nanotechnology* **2019**, *14*, 313.
[21]    D. Alcaraz Iranzo, S. Nanot, E. J. C. Dias, I. Epstein, C. Peng, D. K. Efetov, M. B. Lundeberg, R. Parret, J. Osmond, J.-Y. Hong, J. Kong, D. R. Englund, N. M. R. Peres, F. H. L. Koppens, *Science* **2018**, *360*, 291.
[22]    M. R. Anderson, *Master's thesis,* MISSOURI UNIV-ROLLA, Jan, **2000**.
[23]    D. Melnikau, D. Savateeva, A. Chuvilin, R. Hillenbrand, Y. P. Rakovich, *Opt. Express* **2011**, *19*, 22280.
[24]    A. Cacciola, C. Triolo, O. Di Stefano, A. Genco, M. Mazzeo, R. Saija, S. Patanè, S. Savasta, *ACS Photonics* **2015**, *2*, 971.
[25]    T. H. H. Le, T. Tanaka, *ACS Nano* **2017**, *11*, 9780.
[26]    M. M. Qazilbash, M. Brehm, B.-G. Chae, P.-C. Ho, G. O. Andreev, B.-J. Kim, S. J. Yun, A. Balatsky, M. Maple, F. J. S. Keilmann, *Science* **2007**, *318*, 1750.
[27]    M. Currie, M. A. Mastro, V. D. J. O. M. E. Wheeler, *Optical Materials Express* **2017**, *7*, 1697.
[28]    C. Wan, Z. Zhang, D. Woolf, C. M. Hessel, J. Rensberg, J. M. Hensley, Y. Xiao, A. Shahsafi, J. Salman, S. Richter, Y. Sun, M. Mumtaz Qazilbash, R. Schmidt-Grund, C. Ronning, S. Ramanathan, M. A. Kats, in *arXiv preprint arXiv:1901.02517*, **2019**.
[29]    S.-J. Kim, H. Yun, K. Park, J. Hong, J.-G. Yun, K. Lee, J. Kim, S. J. Jeong, S.-E. Mun, J. J. S. r. Sung, *Scientific reports* **2017**, *7*, 43723.
[30]    T. G. Folland, A. Fali, S. T. White, J. R. Matson, S. Liu, N. A. Aghamiri, J. H. Edgar, R. F. Haglund, Y. Abate, J. D. J. N. c. Caldwell, *Nature communications* **2018**, *9*, 4371.
[31]    D. Wang, Y. R. Koh, Z. A. Kudyshev, K. Maize, A. V. Kildishev, A. Boltasseva, V. M. Shalaev, A. Shakouri, *Nano Letters* **2019**, *19*, 3796.
[32]    O. L. Muskens, L. Bergamini, Y. Wang, J. M. Gaskell, N. Zabala, C. De Groot, D. W. Sheel, J. J. L. S. Aizpurua, Applications, *Light: Science & Applications* **2016**, *5*, e16173.

# Supporting Information

## Qualitative analytical Model of the dual-band antenna

*1. The eigenfrequency and eigenmode of a single disk*

Consider a single gold disk of radius $R$ and height $h$ as in Figure 1a. We choose to approximate it by an oblate spheroid with the major and minor semi-axes being equal to $R$ and $h/2$, correspondingly, see Figure S1a. In the limiting case $h \to 0$, the oblate spheroid becomes an infinitesimally thin disk.

To simplify our analysis of the eigenvalue problem of plasmonic mode excitation in the oblate spheroid, we consider the *quasistatic approximation* assuming that the spheroid is small compared to the resonant wavelength of the mode, *i.e.* $2\pi R/\lambda_r \ll 1$. In this approximation, we neglect the retardation effect due to the finite speed of electromagnetic wave propagation. Then, the plasmonic mode is governed by the Laplace equation and the solution in oblate spheroidal coordinates can be found in[i]. For convenience, we present here a solution for the dipolar mode of interest: The electric potential reads:

$$\varphi = P_2^1(\eta)P_2^1(i\xi)Q_2^1(i\xi_0)e^{i\phi}, \ \xi < \xi_0; \qquad (S1a)$$

$$\varphi = P_2^1(\eta)P_2^1(i\xi_0)Q_2^1(i\xi)e^{i\phi}, \ \xi > \xi_0. \qquad (S1b)$$

Here, $P_2^1(x)$ and $Q_2^1(x)$ are the associated Legendre functions of the second kind. The Cartesian coordinates $(x, y, z)$ are related to the oblate spheroidal coordinates $(\xi, \eta, \phi)$ by[ii]

$$x = f\sqrt{(1-\eta^2)(1+\xi^2)} \cos\phi,$$
$$y = f\sqrt{(1-\eta^2)(1+\xi^2)} \sin\phi,$$
$$z = f\xi\eta.$$

Here, $f$ is the distance between the focal points of the ellipse forming the spheroid. The oblate spheroidal coordinates change in the range: $0 < \xi < \infty$, $-1 < \eta < 1$, $0 < \phi < 2\pi$. The coordinate surfaces $\xi = const$ and $\eta = const$ form ellipsoids and half-hyperboloids of revolution, respectively, see Figure S1b. The $z$-axis is the axis of revolution. The surface $\xi = \xi_0$ corresponds to the surface of the oblate spheroid in question with $\xi_0$ being defined as

$$\xi_0 = \frac{h/2}{\sqrt{R^2 - h^2/4}}.$$

To understand better the spatial structure of the plasmonic mode, it is useful to consider the surface charge density depicted in Figure S1a. This quantity can be calculated from the potential (S1) as

$$\sigma = \frac{\varepsilon_d - 1}{4\pi} \mathbf{n} \cdot \mathbf{E} = \frac{\varepsilon_d - 1}{4\pi} E_\xi = \frac{\varepsilon_d - 1}{4\pi} h_\xi \left.\frac{\partial \varphi}{\partial \xi}\right|_{\xi = \xi_0}, \quad (S2)$$

where $h_\xi = \sqrt{(\xi^2 + \eta^2)/(1 + \xi^2)}$ is the Lamé coefficient and $\varepsilon_d$ is the dielectric permittivity of the spheroid. For a very oblate spheroid, $\xi_0 \ll 1$, $E_\xi$ is approximately equal to the Cartesian $z$-component of the electric field, i.e. $E_\xi \approx E_z$.

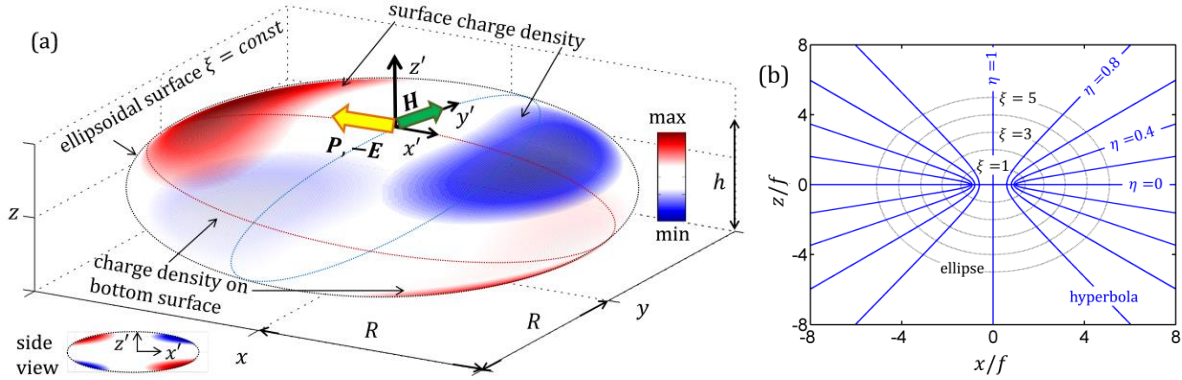

Figure S1. Panel a: surface charge density of the dipolar plasmonic mode of an oblate spheroidal. The electric dipole, electric and magnetic fields at the origin of the local coordinate system $(x', y', z')$ are represented by $\mathbf{P}, \mathbf{E}$ and $\mathbf{H}$, respectively. Panel b: coordinate lines of oblate spheroidal coordinates $\xi = const$ and $\eta = const$ in the Cartesian coordinate system $(x, z)$.

On each face of the spheroid, the charge density has a maximum and a minimum separated by the distance being approximately equal to $R$. Hence, the charge density forms a distributed electric dipole density with the strength $P \approx |\sigma| R / 2$. On the other face, the positions of the maximum and minimum are swapped so that the orientation of the dipole is reversed. At the origin of the local Cartesian coordinate system $(x', y', z')$, see Figure S1a (the global Cartesian system is $(x, y, z)$), the electric field $\mathbf{E}$ is simply oriented along the $x'$-axis whereas the electric dipole is pointing in the opposite direction. The magnetic field $\mathbf{H}$ is orthogonal to $\mathbf{P}$ and $\mathbf{E}$, and directed along the $y'$-axis. This type of the distribution of the surface charge density and electric field is termed the dipolar mode. Note that the quasistatic result (S1) accurately captures the charge and field distributions. cf. Figure S1a and Figure 1b.

The dispersion equation of the dipolar mode of an oblate spheroid reads [i]:

$$\frac{\varepsilon_d(\omega)}{\varepsilon_e(\omega)} = \frac{\left(Q_2^1(i\xi_0)\right)' P_2^1(i\xi_0)}{\left(P_2^1(i\xi_0)\right)' Q_2^1(i\xi_0)} \quad (S3)$$

where $\varepsilon_d(\omega)$ and $\varepsilon_e(\omega)$ are the dielectric permittivities of the (gold) spheroid and the surrounding environment, respectively; the prime stands for the derivative with respect to $\xi_0$. The solution to the Equation (S3) gives resonant frequencies (eigenfrequencies) of the dipolar mode. Using the Drude model for $\varepsilon_d(\omega)$, $\varepsilon_d(\omega) = \varepsilon_0 - \omega_p^2/\omega^2$, the resonant frequency of the dipolar mode can be written as

$$\omega_r = \frac{\omega_p}{\sqrt{\varepsilon_0 + |\text{RHS}|\,\varepsilon_e}}, \tag{S4}$$

where RHS is the right-hand side of Equation (S3). Recall that all results of this section are derived in the quasistatic approximation $2\pi R/\lambda_r \ll 1$ and the smaller the nanoparticle, the higher the accuracy is. For our geometrical parameters, the free-space resonant wavelength is 553 nm according to (S4) and it is 760 nm in full-wave simulations. Despite some discrepancy because of large $R$, (S4) can be used as a starting point in numerical simulations.

*1.2. Excitation of the double-disk resonator*

Consider two gold disks that share the same symmetry axis (oriented along the $z$-axis) forming a vertical dimer. We will mostly be interested in the resonant frequencies of the dimer. Let an external electromagnetic wave with the electric and magnetic vectors $\mathbf{E_0}, \mathbf{H_0}$ excite resonantly only the dipolar mode analyzed earlier. In our simplified consideration, we assume that the vertical separation between the disks $d$ is sufficiently large such that the plasmonic modes of the disks are weakly coupled through their *dipole fields*. From the potential (S1) it follows that the field distribution of the dipolar mode of a single disk, $\mathbf{E_d}\,\mathbf{H_d}$, decays to zero at the distance equal to $R$. Hence, the model presented below is applicable for $d > R$. Then, the electromagnetic field distribution of the dimer can be approximated by the sum of the field distributions of the individual disks. The amplitudes of the fields excited in the 1st and 2nd disks, $V_1$ and $V_2$, are governed by the excitation equations[iii]:

$$\frac{\partial^2 V_1}{\partial t^2} + \omega_r^2 V_1 = -\frac{1}{2W_d} \iint \frac{\partial}{\partial t}(\mathbf{i_{01}} + \mathbf{i_{12}}) \cdot \mathbf{E_d}\, dS, \tag{S5a}$$

$$\frac{\partial^2 V_2}{\partial t^2} + \omega_r^2 V_2 = -\frac{1}{2W_d} \iint \frac{\partial}{\partial t}(\mathbf{i_{02}} + \mathbf{i_{21}}) \cdot \mathbf{E_d}\, dS. \tag{S5b}$$

Here, $\mathbf{i_{01}}$ and $\mathbf{i_{02}}$ are the surface currents on the 1st and 2nd disks, respectively, due to the external excitation field $\mathbf{E_0}, \mathbf{H_0}$; $\mathbf{i_{12}}$ is the surface current on the 1st disk due to the *dipole field* of the 2nd disk whereas $\mathbf{i_{21}}$ is the surface current on the 2nd disk due to the 1st one; $W = \varepsilon/8\pi \int_V E_d^2 dV'$ is the electromagnetic energy of the mode and the integral is taken over the whole volume occupied by the mode. The disk, which is further away from the source by $d$, will be excited with a phase delay of $kd$, i.e. $\mathbf{i_{02}} \propto e^{ikd}\mathbf{i_{01}}$, where $k = 2\pi/\lambda$ is the wavenumber. The excitation equation (S5) has originally the source term in the form[3] $\frac{1}{2W_d} \iiint \frac{\partial}{\partial t}\mathbf{j} \cdot \mathbf{E_d}\, dV$, where j is the current density. Since the field distribution in the disk (spheroid) is approximately constant along the coordinate normal to the surface, we take one integral and reduce the source term to the surface integral in (S5).

The surface current can be expressed through the magnetic field exciting it as:

$$\mathbf{i} = \frac{c}{4\pi}\,\mathbf{n}\times\mathbf{H},$$

where $\mathbf{n}$ is the normal to the surface. For a plane wave $\mathbf{H} = \mathbf{n}\times\mathbf{E}$, thus $\mathbf{i} = -c/(4\pi)\mathbf{E}$. In the near-field zone, the magnetic field of a dipole with the strength $\mathbf{P}$ reads[iv]:

$$\mathbf{H} = \frac{1}{4\pi r^2}\,\mathbf{n}\times\dot{\mathbf{P}},$$

where $r$ is the distance from the source to the observer. After some algebra, one finds that $\mathbf{i} = -c/(16\pi^2 r^2)\dot{\mathbf{P}}$, which allows one to calculate the impact of one disk on another.

We consider excitation by a plane, s-polarized wave with the electric vector being oriented along the x-axis. Then, the excitation equations take on the following form:

$$\frac{\partial^2 V_1}{\partial t^2} + \omega_r^2 V_1 = \frac{c}{8\pi}\frac{1}{W_d}\iint \dot{E}_{0,x} E_{d,x}\, dS + \ddot{V}_2 \frac{Rc}{16\pi^2}\frac{\varepsilon_d - 1}{16\pi}\frac{1}{W_d}\iint \frac{|E_{d,z}(x',y')|E_{d,x}(x,y)}{|\mathbf{r}-\mathbf{r}'|^2}\, dSdS', \quad \text{(S6a)}$$

$$\frac{\partial^2 V_2}{\partial t^2} + \omega_r^2 V_2 = \frac{c}{8\pi}\frac{e^{ikd}}{W_d}\iint g\dot{E}_{0,x} E_{d,x}\, dS + \ddot{V}_1 \frac{Rc}{16\pi^2}\frac{\varepsilon_d - 1}{16\pi}\frac{1}{W_d}\iint \frac{|E_{d,z}(x',y')|E_{d,x}(x,y)}{|\mathbf{r}-\mathbf{r}'|^2}\, dSdS'. \quad \text{(S6b)}$$

Here, the subscripts $x$ and $z$ are used to denote the $x$- and $z$-components of the vector; $g$ is the phenomenological factor that accounts for screening of the incident field by another disk. The second term in the right-hand side of (S6a) and (S6b) contains the double surface integral which accounts for all elementary dipoles. The integral $\iint (1/|\mathbf{r}-\mathbf{r}'|^2)|E_{d,z}|E_{d,x}\, dSdS'$ depends only on the distance between the disks while the transverse coordinates are clearly integrated out. This type of the integral often appears in beam physics and while the exact calculation of it is not straightforward, it turns out that the result of integration can be approximated by a simple function[v] $B(1 + d/w)^{-2}$. The parameters $B$ and $w$ can be found by fitting this function to the result of numerical integration. The physical meaning of $w$ is the characteristic size of the region in which the field distributions $E_{d,x}$ and $E_{d,z}$ overlap. From the surface charge distribution in Figure S1a, one can realize that $E_{d,x}$ attains its maximum in between the extrema of the charge whereas the maximum and minimum of $E_{d,z}$ correspond to those of the charge density. Clearly, the overlap between $E_{d,x}$ and $E_{d,z}$ is small and very roughly an order of magnitude smaller than the disk (spheroid) radius. The parameter $B$ can be thought of as the electrostatic energy of interaction of the physical surface charge density $\sigma(x,y)$ and a virtual charge density $\sigma'(x,y) = E_{d,x}/4\pi$.

*1.3. Double-disk resonator as a system of coupled oscillators*

By introducing auxiliary notations, we can rewrite (S6a) and (S6b) in the form:

$$\frac{\partial^2 V_1}{\partial t^2} + \omega_r^2 V_1 = \alpha\omega_r^2 + \beta\ddot{V}_2, \quad \text{(S7a)}$$

$$\frac{\partial^2 V_2}{\partial t^2} + \omega_r^2 V_2 = e^{ikd}\alpha\omega_r^2 + \beta\ddot{V}_1. \quad \text{(S7b)}$$

The values of $\alpha$ and $\beta$ can easily be found by comparing (S7) and (S6). The set of Equations (S7a) and (S7b) describe a well-known model of two coupled oscillators and we seek a solution in the form $V \propto e^{i\omega t}$. Clearly, in a zero order approximation in $\beta$, $\beta = 0$, the most efficient excitation occurs at the resonant frequency, $\omega = \omega_r$, so that $V \propto e^{i\omega_r t}$. For $\beta \neq 0$, the eigenfrequencies of the vertical dimer are:

$$\omega_\pm = \frac{\omega_r}{\sqrt{1 \mp \beta}} \approx \omega_r \left(1 \pm \frac{\beta}{2}\right). \tag{S8}$$

The separation between the eigenfrequencies increases with the coupling $\beta$ as $\beta\omega_r$ – a well-known effect in the theory of oscillations. The corresponding free-space wavelength ($\lambda = 2\pi c/\omega$) is $\lambda_\mp = 2\pi c/\omega_\pm = \lambda_r \sqrt{1 \mp \beta}$. Recalling that in (S7) the coupling $\beta \propto (1 + d/w)^{-2}$, we see that *the separation between the eigenfrequencies increases as the distance between the disks $d$ decreases*. This supports the results of numerical simulations for $d > R$ shown in Figure 1d. Finally, we re-write the expression for the resonant wavelengths in a convenient parametric form:

$$\lambda_\mp = \lambda_r \left(1 \pm \frac{a}{(1+d/w)^2}\right). \tag{S9}$$

Recall that the result (S9) has been derived under the approximation $d > R$ (the separation between the disks is larger than the disk radius), *i.e.* the disks are coupled through their dipole fields whereas the field distributions of the plasmonic modes do not overlap. At the same time, the comparison with the numerical simulations, see Fig. 1d, showed that the *functional dependence* on $d$ in the Equation (S9) is valid even for $d < R$. Then, by fitting the result (S9) to the corresponding result of our numerical simulations, we found the values $a$ and $w$, which are *valid for all $d$*. It turned out that $a = 4$ nm and $w = 8.33$ nm. This allows using the Equation (S9) for quick analytical estimates for values of $d$, for example, outside the simulation region presented in Fig. 1d. Moreover, the result (S9) can be used for an *order-of-magnitude* estimates for disks of different radii since $w \sim R/10$ (see the discussion about $w$ after (S6)).

**References**

[i] V.V. Klimov, *Nanoplasmonics*, Pan Standford, USA 2014.
[ii] G. Korn, T.Korn, *Mathematical handbook for scientists and engineers*, cMgraw-hill Book Company, New York, USA, 1961.
[iii] R. Collin, *Foundations for microwave engineering,* Wiley-IEEE, USA, 2001.
[iv] J. Jackson, *Classical Electrodynamics*, Wiley, USA, 1999.
[v] G. Shamuilov, A. Mak, K. Pepitone, V. Goryashko. *Appl. Phys. Lett.* **2018**, 113, 204103.